\documentclass[3p]{elsarticle}

\usepackage{hyperref}
\usepackage[utf8]{inputenc}
\usepackage{amsmath}

\graphicspath{{./images/},{./}}

\journal{Journal of \LaTeX\ Templates}

\renewcommand{\vec}{\mathbf} 
\newcommand{\eps}{\varepsilon} 
\newcommand{\id}{\mathbf{1}_{3\times 3}} 
\newcommand{\db}[1]{\overline{\overline{#1}}} 
\newcommand{\e}[1]{\text{e}^{#1}} 

\begin{document}

\begin{frontmatter}


\title{First Principles Homogenization of Periodic Metamaterials and Application to Wire Media}

\author[ITCoimbra]{Sylvain Lanneb\`{e}re}
\author[ITCoimbra]{Tiago A. Morgado}


\author[IST,ITCoimbra]{M\'{a}rio G. Silveirinha\corref{mycorrespondingauthor}}
\cortext[mycorrespondingauthor]{Corresponding author}
\ead{mario.silveirinha@co.it.pt}

\address[ITCoimbra]{Department of Electrical Engineering, University of Coimbra and Instituto de Telecomunica\c{c}\~{o}es, 3030-290 Coimbra, Portugal}
\address[IST]{University of Lisbon -- Instituto Superior T\'ecnico, Department of Electrical Engineering, 1049-001 Lisboa, Portugal}

\begin{abstract}
Here, we present an overview of a first principles homogenization
theory of periodic metamaterials. It is shown that in a rather
general context it is possible to formally introduce effective
parameters that describe the time evolution of macroscopic
(slowly-varying in space) initial states of the electromagnetic
field using an effective medium formalism. The theory is applied to
different types of ``wire metamaterials" characterized by a strong
spatial dispersion in the long wavelength limit. It is highlighted
that the spatial dispersion may tailor in unique ways the wave
phenomena in wire metamaterials leading to exotic tunneling effects
and broadband lossless anomalous dispersion.
\end{abstract}


\end{frontmatter}


\section{Introduction}
 
The interactions between waves and matter play a fundamental role in
most physical processes. It is usually rather challenging to
characterize exactly the wave propagation in macroscopic systems
formed by a large number of identical elements, e.g., in periodic or
random composite materials, due to the complexity of the wave
phenomena at the microscopic level. Fortunately, in many instances,
the detailed microscopic behavior of a wave is of very limited
practical interest. Instead, one can resort to effective medium
theories that provide a simplified description of the wave phenomena
in terms of a limited set of parameters. Effective medium theories
are particularly successful when the wavelength is large with
respect to the characteristic spatial period of the composite
material. In this case, the material may be regarded as a continuum,
and the homogenization
formalism gives a simplified and insightful picture of the wave propagation. 

Effective medium theories have a long history
\cite{sihvola_mixing_2009}. In the case, of light waves the concepts
of ``permittivity" and ``permeability" of a material are as old as
the electromagnetism itself. Similarly, in semiconductor theory the
effects of a periodic electrostatic potential associated with the
ionic lattice can be modeled by an effective electron mass
\cite{kittel_introduction_2004}. In the last two decades, the
interest in effective medium theories has been renewed by the
emergence of the field of metamaterials \cite{smith_homogenization_2006,sjoberg_dispersive_2006,simovski_bloch_2007,simovski_local_2007,silveirinha_metamaterial_2007,silveirinha_generalized_2007,ortiz_effective_2009,silveirinha_nonlocal_2009,costa_finite-difference_2009,smith_analytic_2010,simovski_electromagnetic_2010,silveirinha_time_2011,fietz_current-driven_2010,chebykin_nonlocal_2011,chebykin_nonlocal_2012,alu_first_2011,yaghjian_homogenization_2013,yaghjian_anisotropic_2014,sozio_generalized_2015,simovski_composite_2018,cioranescu_strange_1997,felbacq_homogenization_1997,poulton_noncommuting_2001,poulton_noncommuting_2004,zhikov_spectrum_2005,maurel_sensitivity_2018}. Metamaterials are composite
media formed by properly shaped dielectric or metallic inclusions
embedded in a host medium, which are designed to exhibit
extraordinary behavior such as a negative index of refraction
\cite{pendry_negative_2000}, subwavelength imaging
\cite{luo_subwavelength_2003,belov_subwavelength_2006} or other
applications \cite{capolino_applications_2009}. Usually, in
metamaterials the radiation wavelength $\lambda$ is only moderately
larger than the lattice constant $a$, typically 5-10 times. This
contrasts with natural media where the ratio, $\lambda/a$, is
several orders of magnitude larger than that value, even at optical
frequencies. This property imposes restrictions on the application
of classical homogenization theories to artificial materials
\cite{simovski_bloch_2007,silveirinha_metamaterial_2007,alu_first_2011,yaghjian_homogenization_2013} due to the emergence of spatial
dispersion.

In a spatially dispersive material the electric displacement vector
in a given point of space cannot be written exclusively in terms of
the macroscopic electric field at the same point, but ultimately may
depend on the distribution of the electric field in a neighborhood
that encompasses many unit cells \cite{landau_electrodynamics_1984}.
This non-locality of the electromagnetic response has many important
and nontrivial repercussions on the physical properties of a
material \cite{agranovich_crystal_1984}.

The objective of this review article is to present an up to date
comprehensive description of a general homogenization procedure
first developed in the context of electromagnetic metamaterials
\cite{silveirinha_metamaterial_2007} and later generalized to
semiconductor superlattices \cite{silveirinha_effective_2012}. The
effective medium theory is applicable to a wide range of periodic
physical systems and takes into account both spatial and frequency
dispersion
\cite{silveirinha_nonlocal_2009,silveirinha_effective_2017}. We
illustrate the application of the formalism to ``wire media". This
class of metamaterials is particularly interesting, not only because
it allows for an analytic treatment that describes almost exactly
the actual microscopic response of the metamaterial, but also
because of the richness of the wave phenomena it enables.

The review article is organized as follows: in section
\ref{sec:effective_medium} we describe the general homogenization
scheme of Ref.\cite{silveirinha_effective_2012} that uses as a
starting point a time-domain perspective. In section
\ref{sec:electromagnetic_homogenization}, we focus our analysis on
nonmagnetic and periodic electromagnetic metamaterials and explain
how to find the effective response in the frequency domain. The
homogenization approach is applied to wire media in section
\ref{sec:apps_wire_media}. The nonlocal effective models for
different wire medium topologies are presented in section
\ref{sec:non_local_model}. In section \ref{sec:quasi_static_model},
it is shown that the nonlocality of wire metamaterials emerges
naturally from a quasi-static model with additional state variables
that describe the internal degrees of freedom of the metamaterial.
Some subtleties arising from the nonlocality of the electromagnetic
response, such as the definition of the Poynting vector and the need
for Additional Boundary Conditions (ABCs) are discussed in sections
\ref{sec:Poynting_vector} and \ref{sec:ABC}, respectively. Finally,
in Sec. \ref{sec:anomalous_light_refraction_tunneling} we describe
some exotic wave phenomena due to the spatial dispersion in two
distinct wire medium configurations.

\section{Effective medium theory} \label{sec:effective_medium}

In this section, we present the fundamentals of the homogenization
method originally developed in Refs.
\cite{silveirinha_metamaterial_2007,
silveirinha_effective_2012,silveirinha_effective_2017}. We adopt the
general perspective of Refs. \cite{
silveirinha_effective_2012,silveirinha_effective_2017} where the
effective medium parameters are defined in such a way that they
describe exactly the time-evolution of any macroscopic
(slowly-varying in space) initial wave packet.


\subsection{Microscopic theory} \label{sec:microscopic_schrodinger}
We consider a generic periodic in space physical system whose dynamics is
characterized by a one-body Schr\"{o}dinger-type equation of the
form:
\begin{equation} \label{E:Schrodinger_eq}
\hat{H}\psi= i\hbar\frac{\partial}{\partial t} \psi.
\end{equation}
Here, $\hat{H}$ is the operator that determines the time evolution
of the system and $\psi$ is the state-vector that describes the
state of the system. In general $\psi$ is a multi-component vector
(a spinor). Evidently, this type of formulation is suitable to
characterize the propagation of electron waves in a bulk
semiconductor or in semiconductor or graphene superlattices, and in
such a context $\hat{H}$ is the system Hamiltonian, $\psi$ is the
wave function and $\hbar$ is the reduced Planck constant
\cite{silveirinha_effective_2012,silveirinha_metamaterial-inspired_2012,silveirinha_giant_2014,lannebere_effective_2015}.

Importantly, the propagation of light can also be described using a
similar formulation. Indeed, the Maxwell's equations can be written
in a compact form as \cite{silveirinha_effective_2012}
\begin{align}\label{E:microscopic_Maxwell_eqs_6vector}
\begin{pmatrix} 0  & i \nabla \times \id  \\ -i \nabla \times \id & 0    \end{pmatrix} \cdot \vec{f}
= i \frac{\partial \vec{g}}{\partial t},
\end{align}
where $\vec{f}=\left(  \vec{e}   \quad   \vec{h} \right)^T$ is a
six-element vector with components determined by the microscopic
electric and magnetic fields and $\vec{g}=\left(  \vec{d} \quad
\vec{b} \right)^T$ is a six-element vector with components
determined by the electric displacement and the magnetic induction
fields. In electromagnetic metamaterials the $\vec{f}$ and $\vec{g}$
fields are related by a space-dependent material matrix
$\vec{M}=\vec{M}(\vec{r})$ through the constitutive relation
$\vec{g}=\vec{M}\cdot \vec{f}$. In conventional isotropic media the
material matrix is simply:
\begin{align}
\vec{M}=\begin{pmatrix} \eps \id  & 0  \\ 0 & \mu \id     \end{pmatrix},
\end{align}
where $\eps$ and $\mu$ are the permittivity and permeability,
respectively. Hence, by defining $\hat{H}$ as:
\begin{align}\label{E:H_electromag}
\hat{H}=\hbar\begin{pmatrix} 0  & i \nabla \times \id  \\ -i \nabla \times \id & 0    \end{pmatrix} \cdot \vec{M}^{-1}
\end{align}
and identifying the state vector  with the $\vec{g}$ field, $
\psi=\vec{g}$, the Maxwell's equations can be expressed as in Eq.
\eqref{E:Schrodinger_eq}. It should be noted that in the
electromagnetic case $\hat{H}$ is unrelated to the energy of the
system, and should be simply regarded as an operator that describes
the time evolution of the classical electromagnetic field. Moreover,
in the previous discussion it is implicit that the relevant
materials are nondispersive, i.e., the permittivity $\eps$ and the
permeability $\mu$ are frequency independent. Yet, the formalism can
be generalized to dispersive media, as it is always possible to get
rid of the material dispersion with additional variables \cite{gralak_macroscopic_2010,silveirinha_topological_2018,silveirinha_modal_2019}.
For lossy media, the  $\hat{H}$ operator is non-Hermitian.


\subsection{Spatial averaging and the envelope function} \label{sec:spatial_average}

The envelope function is intuitively the slowly varying part, in
space, of the state vector $\psi$ . It is defined here as:
\begin{align}
 \Psi(\vec{r},t) \equiv \left\{ \psi(\vec{r},t)\right\}_\text{av},
\end{align}
where $\left\{ \right\}_\text{av}$ is a linear operator that performs a spatial averaging. The averaging
operator is completely determined by the response to plane waves, characterized by the function $F(\vec{k})$ such that
\begin{align}
\left\{ \e{i\vec{k}\cdot\vec{r}} \right\}_\text{av} = F(\vec{k}) \e{i\vec{k}\cdot\vec{r}}.
\end{align}
Thus, the action of the averaging operator on a generic plane wave
with wave vector $\vec{k}$ yields another plane wave with the same
wave vector, but with a different amplitude given by $F(\vec{k})$.
Because of the linearity of the operator $\left\{
\right\}_\text{av}$, its action on a generic function is determined
by Fourier theory and is given by a spatial convolution. The
envelope function can be explicitly written as:
\begin{align}
\Psi(\vec{r},t)  = \int d^N\vec{r}'f(\vec{r}')  \psi(\vec{r}-\vec{r}',t),
\end{align}
where $N$ is the space dimension (e.g., $N=3$ for a
three-dimensional metamaterial). The weight function $f$ is the
inverse Fourier transform of $F$ so that:
\begin{align}
f(\vec{r})  = \frac{1}{(2\pi)^N}\int d^N\vec{k} ~F(\vec{k})  \e{i\vec{k}\cdot\vec{r}}.
\end{align}
Related ideas have been developed by Russakov in the context of
macroscopic electromagnetism \cite{russakoff_derivation_1970}. It is assumed that the averaging operator corresponds to an ideal
low pass spatial filter such that:
\begin{align}F(\vec{k})=
\begin{cases}
1, & \vec{k} \in \text{B.Z.}\\ 0, & \text{otherwise}
\end{cases}.
\end{align}
In this article the set B.Z. stands for the first Brillouin zone of
the periodic lattice, but sometimes other choices can be relevant
\cite{silveirinha_effective_2017}. The envelope function
$\Psi(\vec{r},t)$ has no significant spatial fluctuations on the
scale of a unit cell, i.e., the microscopic fluctuations are
filtered out by the averaging operator. Hence, $\Psi(\vec{r},t)$
determines the macroscopic state vector. In general, we say that a
given state vector $\psi$ is macroscopic when it stays invariant
under the operation of spatial averaging:
\begin{align}
\psi(\vec{r})  = \left\{ \psi(\vec{r})   \right\}_\text{av}, \quad \text{(macroscopic state vector)}.
\end{align}
Importantly, a macroscopic state cannot be more localized in space than the characteristic period of the material.

\subsection{The effective Hamiltonian} \label{sec:effective_hamiltonian}

\begin{figure}[!ht]
\centering
\includegraphics[width=.55\linewidth]{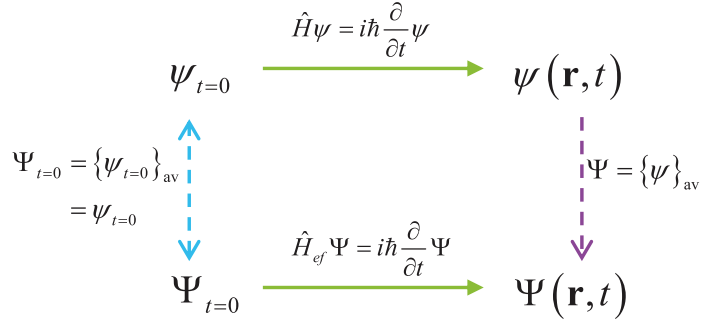}
         \caption{Schematic relation between the time evolutions determined by the macroscopic and
microscopic Hamiltonians: for an initial macroscopic state the effective medium formulation
ensures that $\Psi=\left\{ \psi\right\}_\text{av}$ for $t>0$. Reprinted with permission \cite{silveirinha_metamaterial-inspired_2012}.}
         \label{fig:effective_medium_principle}
\end{figure}

%
The effective Hamiltonian is the operator that describes the time
evolution of the envelope function. Specifically, suppose that the
initial state vector is macroscopic, so that $\psi_{t=0}=\Psi_{t=0}$
. In general, the time evolution of an initial macroscopic state
does not yield a macroscopic state at a later time instant, i.e.,
$\psi(\vec{r},t)\neq \Psi(\vec{r},t)$ for $t>0$. We define the
effective Hamiltonian $\hat{H}_\text{ef}$ such that
$\Psi(\vec{r},t)$ calculated using $\hat{H}_\text{ef}$ is coincident
with the spatially-averaged microscopic state vector
$\left\{\psi(\vec{r},t)\right\}_\text{av}$, where $\psi(\vec{r},t)$
is determined by the microscopic Hamiltonian $\hat{H}$
\cite{silveirinha_effective_2012,silveirinha_metamaterial-inspired_2012}.
These ideas are illustrated in the diagram of Figure
\ref{fig:effective_medium_principle}.

The time evolution of the macroscopic state vector is determined by
a generalized Schr\"{o}dinger equation:
\begin{align}
 \left( \hat{H}_\text{ef} \Psi \right)(\vec{r},t)=i \hbar \frac{\partial}{\partial t}\Psi(\vec{r},t).
\end{align}
From the definition of the effective Hamiltonian it is clear that it must ensure that:
\begin{align}
 \left\{ \hat{H} \psi \right\}_\text{av}=\hat{H}_\text{ef} \Psi.
\end{align}
Because of linearity, the action of the effective Hamiltonian on the
wave function can be expressed as a convolution in space and in time
\cite{silveirinha_effective_2012}:
\begin{align}\label{E:convolution_Hef}
 \left( \hat{H}_\text{ef} \Psi \right)(\vec{r},t)=\int d^N\vec{r}'\int_0^t dt' \vec{h}_\text{ef}(\vec{r}-\vec{r}',t-t') \cdot \Psi(\vec{r}',t').
\end{align}
Note that the kernel $\vec{h}_\text{ef}$ is a function of
$\vec{r}-\vec{r}'$. We shall see below that this possible because
the spatial averaging operation essentially eliminates the spatial
granularity of the system. In general, the kernel
$\vec{h}_\text{ef}(\vec{r},t)$ is represented by a square matrix
$\left[\vec{h}_{\sigma,\sigma'}\right]$ because $\Psi$ is a
multi-component vector. In the photonic case the dimension of
$\vec{h}_\text{ef}$ is $S=6$. Equation \eqref{E:convolution_Hef}
shows that the effective Hamiltonian depends on the past history
($0<t'<t$) and on the surroundings ($\vec{r}'\neq \vec{r}$) of the observation point, rather than just on the instantaneous and
local value of $\Psi$. It is convenient to introduce the Fourier
transform of $\vec{h}_\text{ef}(\vec{r},t)$ defined as:
\begin{align}
\boldsymbol{\mathcal{H}}_\text{ef}(\vec{k},\omega)=\int d^N\vec{r}\int_0^{+\infty} dt~ \vec{h}_\text{ef}(\vec{r},t) \e{i\omega t }\e{-i\vec{k}\cdot\vec{r}}.
\end{align}
The Fourier transform is bilateral in space and unilateral in time. The unilateral
Fourier transform in time can also be regarded as a Laplace transform. In the
Fourier domain, the action of the effective Hamiltonian reduces to a simple
multiplication:
\begin{align}
\left( \hat{H}_\text{ef} \Psi \right)(\vec{k},\omega)=\boldsymbol{\mathcal{H}}_\text{ef}(\vec{k},\omega) \cdot \Psi (\vec{k},\omega).
\end{align}
Here, $\Psi (\vec{k},\omega)$ is the Fourier transform of the
macroscopic state vector,
\begin{align}
\Psi (\vec{k},\omega)=\int d^N\vec{r}\int_0^{+\infty} dt~ \Psi(\vec{r},t) \e{i\omega t }\e{-i\vec{k}\cdot\vec{r}},
\end{align}
and  $\left( \hat{H}_\text{ef} \Psi \right)(\vec{k},\omega)$ is
defined similarly. The integral in $\vec{r}$ is over all space. Note that for now the system is assumed to be periodic and unbounded, so that the effect of boundaries is disregarded. The convergence of the unilateral Fourier
transform is guaranteed in the upper-half frequency plane,
Im$(\omega)>0$.

The function $\boldsymbol{\mathcal{H}}_\text{ef}(\vec{k},\omega)$
completely determines the effective Hamiltonian. Because of the
properties of the spatial averaging operator, it is possible to
enforce that:
\begin{align}
\boldsymbol{\mathcal{H}}_\text{ef}(\vec{k},\omega)=0, \quad \text{when } \vec{k}\notin \mathrm{B.Z..}
\end{align}
This property ensures that the effective Hamiltonian is a smoothened
version of the microscopic Hamiltonian. In the following
subsections, it is explained how
$\boldsymbol{\mathcal{H}}_\text{ef}(\vec{k},\omega)$ can be
calculated for $\vec{k} \in$ B.Z..

\subsubsection{Calculation of $\boldsymbol{\mathcal{H}}_\text{ef}(\vec{k},\omega)$ with a time domain approach}
Let us consider an initial macroscopic state of the form $\psi_{t=0}
\sim \e{i \vec{k}\cdot\vec{r}}\vec{u}_l$ where the wavevector
$\vec{k}$   can take any value in the B.Z.. Here, $(\vec{u}_l)$
represents a basis of unit vectors that generates the
$S$-dimensional vector space wherein $\psi$ is defined. Because of
the periodicity of the system, the microscopic time evolution of
this initial state yields a state vector  $\psi(\vec{r},t)$ with the
Bloch property. In fact, $\psi(\vec{r},t)\e{-i \vec{k}\cdot\vec{r}}$
is a periodic function in space for any fixed $t$. For the same
reason, $\hat{H}\psi$ has also the Bloch property. Crucially, the
operation of spatial averaging only retains spatial harmonics with
wave vector inside the B.Z., and hence it follows that the
dependence of $\left\{  \psi \right\}_\text{av}$  and $\left\{
\hat{H}\psi \right\}_\text{av}$ on the spatial coordinates is of the
form $\e{i \vec{k}\cdot\vec{r}}$ for any time instant. In other
words, within the effective medium approach the time evolution of a
plane wave-type initial state yields another plane wave-type state,
such that the homogenized structure behaves as a continuum.
Moreover, for Bloch modes it is possible to write:
\begin{subequations}\label{E:average_psi_Hpsi}
\begin{align}
\left\{  \psi \right\}_\text{av}(\vec{r},t) &=\psi_\text{av}(t) \cdot \e{i\vec{k}\cdot\vec{r}}, \label{E:average_opetator_Bloch}\\
\left\{  \hat{H} \psi \right\}_\text{av}(\vec{r},t) &= \left( \hat{H}\psi\right)_\text{av}(t) \cdot \e{i\vec{k}\cdot\vec{r}},
\end{align}
\end{subequations}
with
\begin{subequations}\label{E:average_psi_Hpsi_2}
\begin{align}
\psi_\text{av}(t) &= \frac{1}{V_\text{cell}} \int_{\Omega} d^N\vec{r}~ \psi(\vec{r},t) \e{-i\vec{k}\cdot\vec{r}} ,\\
\left( \hat{H}\psi\right)_\text{av}(t) &= \frac{1}{V_\text{cell}} \int_{\Omega} d^N\vec{r}~ \hat{H}\psi(\vec{r},t) \e{-i\vec{k}\cdot\vec{r}} ,
\end{align}
\end{subequations}
where $\Omega$ represents the unit cell and $V_\text{cell}$ is the respective volume.
Taking now into account that $\Psi=\left\{  \psi \right\}_\text{av}$ and $\hat{H}_\text{ef}\Psi=\left\{ \hat{H} \psi \right\}_\text{av}$, and
substituting Eq. \eqref{E:average_psi_Hpsi} into Eq. \eqref{E:convolution_Hef}, it is seen after straightforward manipulations that:
\begin{align}
\left( \hat{H}\psi\right)_\text{av}(\omega) &= \boldsymbol{\mathcal{H}}_\text{ef}(\vec{k},\omega)\cdot \psi_\text{av}(\omega).
\end{align}
In the above, $\psi_\text{av}(\omega)$ and $\left( \hat{H}\psi\right)_\text{av}(\omega)$ stand for the unilateral Fourier (Laplace)
transforms of the functions in Eq. \eqref{E:average_psi_Hpsi_2}. Hence, if we denote $\psi^{(l)}$, $l=1,\dots,S$ as the
microscopic state vector determined by the time evolution of the initial state $\psi_{t=0}^{(l)}= i /\hbar ~\e{i\vec{k}\cdot\vec{r}}\vec{u}_l$ (the proportionality constant was fixed as $i /\hbar$ for convenience), it follows from the previous analysis that the effective Hamiltonian is given by:
\begin{equation}\label{E:effective_Ham_time}
\boldsymbol{\mathcal{H}}_\text{ef}(\vec{k},\omega)=\left[ \left( \hat{H}\psi^{(1)} \right)_\text{av}  \dots \relax \left( \hat{H}\psi^{(S)}  \right)_\text{av}   \right] \cdot \left[ \psi_\text{av}^{(1)} \dots  \psi_\text{av}^{(S)}\right]^{-1}.
\end{equation}
Here $\boldsymbol{\mathcal{H}}_\text{ef}$ and the two objects delimited by the square brackets are $S \times S$ matrices.
Thus, $\boldsymbol{\mathcal{H}}_\text{ef}(\vec{k},\omega)$ can be written as the product of two matrices, whose columns
are determined by the vectors $\psi_\text{av}^{(l)}(\omega)$ and $\left( \hat{H}\psi^{(l)}  \right)_\text{av} (\omega)$.\\
In summary, for an arbitrary $\vec{k}\in$ B.Z. the effective
Hamiltonian can be found by solving $S$ microscopic time evolution
problems associated with initial states of the form
$\psi_{t=0}^{(l)}= i /\hbar ~\e{i\vec{k}\cdot\vec{r}}\vec{u}_l$. The
effective Hamiltonian is written in terms of the Fourier transforms
in time of the functions \eqref{E:average_psi_Hpsi_2}.

\subsubsection{Calculation of $\boldsymbol{\mathcal{H}}_\text{ef}(\vec{k},\omega)$ with a frequency domain approach}
The effective Hamiltonian may also be determined based on frequency
domain calculations. To prove this we note that $\psi_\text{av} (\omega)$ and $\left( \hat{H}\psi   \right)_\text{av} (\omega)$ can be written explicitly as:
\begin{subequations}\label{E:average_psi_Hpsi_2_omega}
\begin{align}
\psi_\text{av}(\omega) &= \frac{1}{V_\text{cell}} \int_{\Omega} d^N\vec{r}~ \psi(\vec{r},\omega) \e{-i\vec{k}\cdot\vec{r}} , \label{E:averaged_bloch_mode_frq}\\
\left( \hat{H}\psi\right)_\text{av}(\omega) &= \frac{1}{V_\text{cell}} \int_{\Omega} d^N\vec{r}~ \hat{H}\psi(\vec{r},\omega) \e{-i\vec{k}\cdot\vec{r}} ,
\end{align}
\end{subequations}
where $\psi(\vec{r},\omega)$ is the unilateral Fourier transform of $\psi(\vec{r},t)$. Applying the unilateral Fourier (Laplace) transform to both members of the microscopic Schr\"odinger equation \eqref{E:Schrodinger_eq} and using the property $\partial_t \psi(\vec{r},t)\leftrightarrow - i \omega \psi(\vec{r},\omega)-\psi_{t=0}(\vec{r})$, it follows that:
\begin{equation} \label{E:Schrodinger_eq_laplace}
\left[ \hat{H} -\hbar \omega \right] \cdot \psi(\vec{r},\omega)= -i\hbar\psi_{t=0}(\vec{r}).
\end{equation}
Hence, $\psi^{(l)}(\vec{r},\omega)$ can be directly found by solving the above equation for $-i\hbar\psi_{t=0}^{(l)}=\e{i\vec{k}\cdot\vec{r}} \vec{u}_l$, with $l=1,\dots,S$. Once $\psi^{(l)}(\vec{r},\omega)$ is known one can determine $\psi_\text{av}^{(l)}$ and $\left( \hat{H}\psi^{(l)}\right)_\text{av}$ using Eq. \eqref{E:average_psi_Hpsi_2_omega}, and finally obtain the effective Hamiltonian from Eq. \eqref{E:effective_Ham_time}.\\
It is interesting to note that for $-i\hbar\psi_{t=0}^{(l)}=\e{i\vec{k}\cdot\vec{r}} \vec{u}_l$ equation \eqref{E:Schrodinger_eq_laplace} implies that $\left( \hat{H}\psi^{(l)}\right)_\text{av}-\hbar \omega \psi_\text{av}^{(l)} = \vec{u}_l$. Substituting this result into Eq. \eqref{E:effective_Ham_time} one may also write the effective Hamiltonian as:
\begin{equation}\label{E:effective_Ham_frq}
\boldsymbol{\mathcal{H}}_\text{ef}(\vec{k},\omega)= \hbar \omega + \left[ \psi_\text{av}^{(1)} \dots  \psi_\text{av}^{(S)}\right]^{-1}.
\end{equation}

\subsection{Stationary states} \label{sec:stationary_states}
The spectrum of the effective Hamiltonian is exactly coincident with
the spectrum of the microscopic Hamiltonian
\cite{silveirinha_effective_2012} (here, for simplicity it is
assumed that there are no ``dark states'', for a discussion see
\cite{silveirinha_effective_2012}). The energy spectrum of the
macroscopic Hamiltonian is determined by the nontrivial solutions of
the stationary Schr\"odinger equation
\begin{align}\label{E:stationary_states_H_ef}
 \left[ \left. \boldsymbol{\mathcal{H}}_\text{ef}(\vec{k},\omega) \right|_{\omega=E/\hbar}-E \right] \cdot \Psi=0,
\end{align}
where $E$ stands for the energy of a certain stationary state. For
example, in the electromagnetic case the photonic band structure
calculated with the effective Hamiltonian is coincident with the
exact band structure obtained using a microscopic theory
\cite{silveirinha_metamaterial_2007}. The enunciated result follows
from the fact that in a time evolution problem (with no source
excitation) the state vector can be written as a superposition of
eigenmodes. The eigenmodes have a time variation of the form
$\e{-i\omega_n t}$, being $\omega_n=E_n/\hbar$ the relevant
eigenfrequencies. Importantly, since the macroscopic and microscopic
state vectors are related by the spatial-averaging operation
($\Psi=\left\{ \psi \right\}_\text{av}$), both $\Psi$ and $\psi$
have the same-type of time oscillations. In other words, the
averaging affects only the space coordinates, while the time
coordinate is not averaged in any manner. As a consequence, the
spectrum of the microscopic and macroscopic Hamiltonians must be the
same. For a detailed mathematical proof of this property the reader is referred to Appendix C of Ref.\cite{silveirinha_effective_2012}.

\section{ The electromagnetic case} \label{sec:electromagnetic_homogenization}
The formalism of the previous section when applied to the
electromagnetic case (Eq. \ref{E:microscopic_Maxwell_eqs_6vector})
yields a 6$\times$6 effective Hamiltonian of the form
\cite{silveirinha_effective_2017}:
\begin{align}
\boldsymbol{\mathcal{H}}_\text{ef}(\vec{k},\omega)=\hbar \begin{pmatrix}0 & - \vec{k}\times \vec{1}_{3\times3}\\ \vec{k}\times \vec{1}_{3\times3} & 0 \end{pmatrix} \cdot \vec{M}^{-1}_\text{ef}(\vec{k},\omega),
\end{align}
where $\vec{M}_\text{ef}(\vec{k},\omega)$ is the effective material
matrix that links the averaged fields $\left\{ \vec{f} \right\}_\text{av}$ and $\left\{ \vec{g} \right\}_\text{av}$ of
\eqref{E:microscopic_Maxwell_eqs_6vector},
\cite{silveirinha_effective_2012,silveirinha_effective_2017}. 
%
%
For metamaterials made of non-magnetic particles the material matrix is
of the form
\begin{align}
\vec{M}_\text{ef}(\vec{k},\omega)=\begin{pmatrix}\db{\eps}_\text{ef}(\vec{k},\omega)&0\\0&\mu_0\vec{1}_{3\times3}\end{pmatrix}.
\end{align}
Thus, the homogenization problem reduces to the determination of the
nonlocal effective permittivity
$\db{\eps}_\text{ef}(\vec{k},\omega)$. The permittivity can be found
using the source-driven homogenization theory developed for
electromagnetic metamaterials \cite{silveirinha_metamaterial_2007}.
As shown in \cite{silveirinha_effective_2017} the effective response
obtained with this theory is exactly coincident with the one
obtained with the general theory of previous section. Below, we
quickly review the main ideas of the source-driven homogenization,
highlighting that the homogenization problem can be reduced to an
integral equation \cite{silveirinha_metamaterial_2007}.

We consider a generic nonmagnetic periodic metamaterial described by
the periodic permittivity $\eps_r(\vec{r},\omega)=\eps_r(\vec{r}+\vec{R},\omega)$
with $\vec{R}$ a vector of the Bravais lattice. Assuming a time
variation of the form $\e{-i\omega t}$, the microscopic Maxwell
equations in this system are
\begin{subequations}\label{E:microscopic_Maxwell_eqs}
\begin{align}
 \nabla \times\vec{e}&=i \omega \vec{b} \\
\nabla \times\frac{\vec{b}}{\mu_0}&=\vec{j}_e - \eps_0\eps_r i\omega \vec{e}
\end{align}
\end{subequations}
where $\vec{e},\vec{b}$ are the microscopic electric and magnetic
field, respectively and $\vec{j}_e$ is an applied (macroscopic)
electric current density that acts as a source of the
electromagnetic fields. The applied current density is assumed to
have the Bloch property and enforces a desired spatial variation
within the unit cell. This means that the pair of parameters
$(\omega,\vec{k})$ characterizing the time and space variations of
the fields are independent of each other and do not need to be
associated with an eigenmode. The applied current plays the same
role as the initial state $\psi_{t=0}$ in the formulation of last
section.

By applying the averaging operator \eqref{E:average_opetator_Bloch}
to the microscopic  Maxwell equations
\eqref{E:microscopic_Maxwell_eqs}, one obtains the macroscopic
Maxwell equations:
\begin{subequations}
\begin{align}
\vec{k} \times\vec{E}_\text{av}&= \omega \vec{B}_\text{av}, \\
\vec{k} \times\frac{\vec{B}_\text{av}}{\mu_0}&=-i\vec{J}_{e,\text{av}} -  \omega \vec{P}_\text{g}   - \eps_0 \omega \vec{E}_\text{av}, \label{E:averaged_Maxwell_eq_2}
\end{align}
\end{subequations}
where $\vec{E}_\text{av},\vec{B}_\text{av}$ and
$\vec{J}_{e,\text{av}}$ are the averaged $\vec{e},\vec{b}$ and
$\vec{j}_e$, respectively, defined according to
Eq.\eqref{E:averaged_bloch_mode_frq}. The averaged induced
polarization $\vec{P}_\text{g}$ is given by
\begin{align}
\frac{\vec{P}_\text{g}}{\eps_0}=\frac{1}{V_\text{cell}} \int_{\Omega} (\eps_r(\vec{r})-1)\vec{e}(\vec{r})\e{-i\vec{k}\cdot\vec{r}} d^3\vec{r} .
\end{align}
For system containing perfectly electric conducting (PEC) surfaces,
the integration over the unit cell volume in the previous expression can
be transformed into a surface integral, see
\cite{silveirinha_metamaterial_2007,silveirinha_nonlocal_2009} for
more details.

The nonlocal effective permittivity is defined through the relation
between the averaged electric field and the averaged induced
polarization:
\begin{align} \label{E:def_effective_permittivity}
 \db{\eps}_\text{ef}(\omega,\vec{k}) \cdot \vec{E}_\text{av}= \eps_0 \vec{E}_\text{av}+\vec{P}_\text{g}.
\end{align}

As shown in \cite{silveirinha_metamaterial_2007}, for every pair
$(\omega,\vec{k})$ the homogenization problem can be reduced to an
integral equation. The unknown of the integral equation is the
microscopic vector field $\vec{p}_\text{ind}(\vec{r})=
(\eps_r(\vec{r})-1)\vec{e}(\vec{r})$ and the excitation is the
averaged electric field $\vec{E}_\text{av}$. A solution of the
problem can be formally constructed using the Method of Moments
(MoM). The unknown $\vec{p}_\text{ind}$ is expanded as
\begin{align}\label{E:expansion_MoM_f}
\vec{p}_\text{ind}= \sum_n c_n \vec{w}_{n,\vec{k}},
\end{align}
where the set of expansion functions $\vec{w}_{n,\vec{k}}$ has the
Bloch property and is assumed to be a complete set in
$\left\{\vec{r}:\eps_r(\vec{r})-1\neq 0\right\}$.

For simplicity, next we focus on the case where the metamaterial
inclusions can be modeled as impedance boundaries, characterized by
some surface impedance $Z_s$ \cite{pozar_microwave_2011}. The
surface impedance links the tangential electric field
$\vec{E}_\text{tan}$ at the boundary surface ($\partial D$) with the
current surface density $\vec{J}_s =
\hat{\boldsymbol{\nu}}\times\vec{H}$, as $\vec{E}_\text{tan}=Z_s
\vec{J}_s$ \cite{silveirinha_artificial_2009}. Here,
$\hat{\boldsymbol{\nu}} $ is the unit normal vector oriented toward
the exterior of the inclusion. A PEC inclusion is described by the
surface impedance $Z_s = 0$. It can be shown that the effective
permittivity is given by
\cite{silveirinha_metamaterial_2007,silveirinha_nonlocal_2009}
\begin{align} \label{E:eps_ef_MoM}
 \frac{\db{\eps}_\text{ef}}{\eps_0}(\omega,\vec{k})= \db{\vec{I}} + \frac{1}{V_\text{cell}}\sum_{m,n} \chi^{m,n}\int_{\partial D} \vec{w}_{m,\vec{k}}(\vec{r})\e{-i\vec{k}\cdot \vec{r}}ds \otimes \int_{\partial D} \vec{w}_{n,-\vec{k}}(\vec{r})\e{i\vec{k}\cdot \vec{r}}ds
\end{align}
%
\begin{multline}\label{E:chi_MoM}
 \chi_{m,n} = \int_{\partial D}\int_{\partial D} \left[ \nabla_s\cdot \vec{w}_{m,-\vec{k}}(\vec{r})\nabla_{s'}\cdot \vec{w}_{n,\vec{k}}(\vec{r}')  - (\omega/c)^2 \vec{w}_{m,-\vec{k}}(\vec{r})\cdot \vec{w}_{n,\vec{k}}(\vec{r}')\right]\Phi_{p0}(\vec{r}|\vec{r}';\omega,\vec{k})
 dsds'\\
-i\omega\eps_0 Z_s \int_{\partial D}
\vec{w}_{m,-\vec{k}}(\vec{r})\cdot \vec{w}_{n,\vec{k}}(\vec{r}) ds.
\end{multline}
In the above $\Phi_{p0}$ is the Green's function introduced in Eq.
(35b) of \cite{silveirinha_metamaterial_2007}, $\nabla_s$ stands for
the surface divergence of a tangential vector field and the matrix
$[\chi^{m,n}]$ is the inverse of $[\chi_{m,n}]$. In the next
section, we illustrate the application of the above formulas to the
case of wire metamaterials.

As shown in
\cite{silveirinha_metamaterial_2007,silveirinha_nonlocal_2009}, Eq. \eqref{E:eps_ef_MoM} can be
generalized to the case of volumetric dielectric inclusions. The MoM
formulation is particularly well suited to characterize the
effective response of metamaterials made of metallic structures. Due
to this reason, for dielectrics it is typically more practical to
solve the homogenization problem with finite differences methods in
the frequency \cite{costa_finite-difference_2009} or in the time
domain \cite{silveirinha_time_2011}.

\section{Application to wire metamaterials} \label{sec:apps_wire_media}
Next, we apply the homogenization method to periodic arrays of thin
metallic wires. Wire metamaterials are generically characterized by
a strong spatial dispersion in the long wavelength limit.

\subsection{Nonlocal effective models}\label{sec:non_local_model}

In the following subsections we obtain the effective medium
responses of three different wire metamaterials: the uniaxial wire
medium, the double wire medium and the 3D connected wire mesh. In
all cases, it will be assumed that the metallic wires are thin, $R
\ll a$, where $R$ is the radius of the wires and $a$ is the spatial
period. The wires are modeled as impedance boundaries characterized
by the surface impedance $Z_s = \frac{2i}{\omega \eps_0(\eps_m-1)R}$
where $\eps_m$ is the metal relative permittivity. The wires are   embedded in a host medium of permittivity $\eps_h$.

\subsubsection{Uniaxial wire medium}\label{sec:uniaxial}
\begin{figure}[!ht]
\centering
\includegraphics[width=\linewidth]{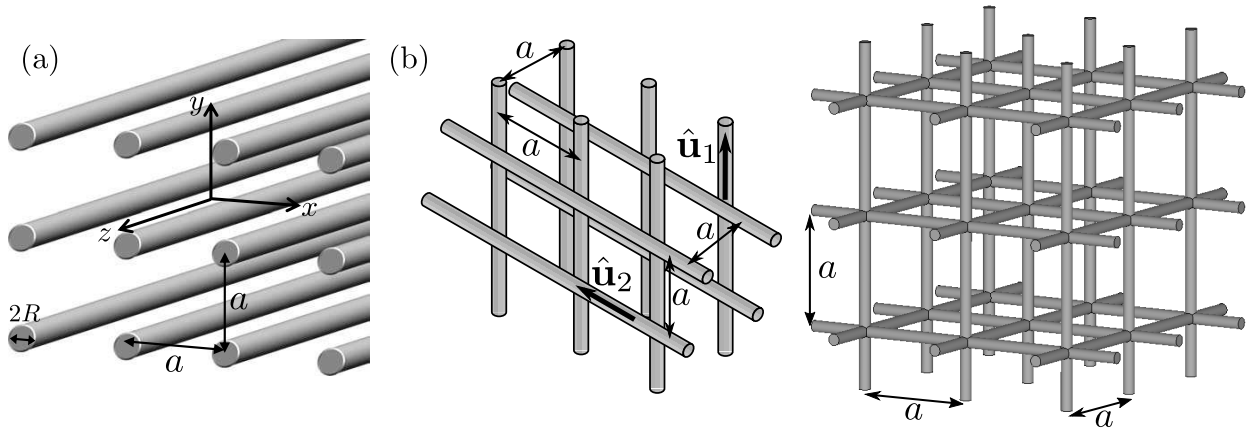}
         \caption{(a) Uniaxial wire medium formed by a square array (period $a$) of infinitely long metallic rods  oriented along the  $\hat{\vec{z}}$ direction. (b) Double wire medium formed by two non-connected arrays of parallel wires arranged in a cubic lattice with lattice constant $a$. The two arrays of wires are oriented along $\hat{\vec{u}}_1$ and $\hat{\vec{u}}_2$ and the distance between adjacent wires is $a/2$. (c) 3D wire mesh formed by a connected array of orthogonal metallic wires. In (a)-(c) the wires radius is $R$ and their permittivity is $\eps_m$.}
         \label{fig:different_wire_meshes}
\end{figure}

The simplest example of a wire metamaterial is the so-called
uniaxial wire medium. It consists of a square lattice of parallel
and infinitely long metallic wires oriented along a fixed direction,
here taken as the  $\hat{\vec{z}}$ direction as represented in Fig.
\ref{fig:different_wire_meshes}(a).
%

The study of such systems has a long history (dating back to the
1950s) that was renewed at the turn of this century after the
discovery of negative index metamaterials
\cite{rotman_plasma_1962,pendry_extremely_1996,maslovski_wire_2002,belov_dispersion_2002,belov_strong_2003,nefedov_wire_2009,simovski_wire_2012}.
%
%

As shown in \cite{silveirinha_artificial_2009}, the application of
the homogenization scheme of section
\ref{sec:electromagnetic_homogenization} to this wire metamaterial
is particularly simple. Indeed, the current density induced on the
metallic wires surface can be accurately modeled by a single
expansion function:
\begin{align}
 \vec{w}_{1,\vec{k}}(\vec{r}) = \frac{\e{i \vec{k} \cdot \vec{r}}}{2\pi R}
 \hat{\vec{z}}.
\end{align}
Note that the electric current density is proportional to
$\vec{p}_\text{ind}$. Using $\vec{w}_{1,\vec{k}}(\vec{r})$ in
equations \eqref{E:eps_ef_MoM} and \eqref{E:chi_MoM}, it can be
shown that the nonlocal effective  permittivity reduces to
\cite{silveirinha_nonlocal_2009,silveirinha_artificial_2009,silveirinha_nonlocal_2006}
\begin{align}\label{E:eps_ef_uniaxial}
\frac{\db{\eps}_\text{ef}}{\eps_0}(\omega,\vec{k})= \db{\vec{I}} +
\frac{1}{\frac{1}{f_V(\eps_m-1)}+\frac{1}{\beta_p^2}\left(k_z^2-\frac{\omega^2}{c^2}\right)
}\hat{\vec{z}} \otimes \hat{\vec{z}},
\end{align}
where $f_V=\pi R^2/a^2$ is the volume fraction of the wires and $\beta_p$ is the
plasma wavenumber for an array of parallel PEC wires. The parameter
$\beta_p$ depends solely on the system geometry (see the next
subsection for the expression and
Ref.\cite{silveirinha_artificial_2009} for further details).

As seen, the effective permittivity of the uniaxial wire medium
depends on the $z$ component of the wavevector along the wires
($k_z$), which leads to a pole of the material response at low
frequencies (for good conductors $\eps_m\to-\infty$ and the pole occurs for $k_z \approx \omega/c$ ). Thereby,
the spatial dispersion effects are rather strong. This feature has
several nontrivial consequences, e.g., it implies that the medium
may support two modes with the same polarization
\cite{belov_strong_2003,silveirinha_nonlocal_2006,nefedov_reflection_2006}. For a full
discussion about the uniaxial wire medium modes the
reader is referred to \cite{silveirinha_nonlocal_2006}. The uniaxial
wire medium has interesting applications in subwavelength imaging
when operated in the  canalization regime
\cite{belov_subwavelength_2006,belov_canalization_2005,belov_resolution_2006,belov_experimental_2006,belov_transmission_2008,ikonen_magnification_2007,shvets_guiding_2007,silveirinha_subwavelength_2007,morgado_transport_2009,morgado_experimental_2010,latioui_near-field_2016}.
%

\subsubsection{Double wire mesh}\label{sec:non_connected_wire_mesh}

A more complex situation from the homogenization perspective occurs
when a second array of parallel wires with a different
orientation is inserted in between the first set of wires (see
Fig.\ref{fig:different_wire_meshes}(b)).
%
Such structures are usually referred to as double wire meshes, and
can have several interesting applications and rather exotic physics
\cite{silveirinha_homogenization_canonical_2005,silveirinha_nonresonant_2008,silveirinha_broadband_2009,silveirinha_anomalous_2009,morgado_ultraconfined_2011,morgado_negative_2012,costa_achromatic_2012,morgado_reversed_2014}.
%
%
While the expression of the nonlocal effective permittivity of this
metamaterial is well known  \cite{silveirinha_nonresonant_2008}, its
direct derivation using the homogenization formalism of Sect.
\ref{sec:electromagnetic_homogenization} was not reported previously
in the literature. Since we believe that proof is pedagogical we do
so in the following.

The wire arrays are oriented along the generic directions
$\hat{\vec{u}}_1$ and $\hat{\vec{u}}_2$. For simplicity we restrict
our analysis to PEC wires ($Z_s= 0$), orthogonal to each other
$\hat{\vec{u}}_m \cdot \hat{\vec{u}}_n=\delta_{m,n}$, with
$m,n=1,2$, and consider a cubic lattice with period $a$. Similar to
the case of the uniaxial wire medium, one expansion function per
wire (two in total) is sufficient to obtain an approximate
analytical expression for the effective permittivity. The expansion
function that models the density of current induced in the $n$th
wire oriented along $\hat{\vec{u}}_n$ (here assumed parallel to one
of the coordinate axes) is taken as
\begin{align} \label{E:expansion_function_nonconnected}
 \vec{w}_{n,\vec{k}}(\vec{r}) = \frac{\e{i \vec{k} \cdot \vec{r} }}{2\pi R}  \hat{\vec{u}}_n, \quad n=1,2.
\end{align}
%
%
%
Substituting the above formula into \eqref{E:eps_ef_MoM}, one finds
that the effective permittivity can be written as:
\begin{align} \label{E:eps_ef_non_connected_1}
 \frac{\db{\eps}_\text{ef}}{\eps_0}(\omega,\vec{k})  &= \db{\vec{I}} + \frac{1}{a} \sum_{m,n} \chi^{m,n}  \hat{\vec{u}}_m    \otimes    \hat{\vec{u}}_n.
\end{align}
To obtain $\chi^{m,n}$, we substitute
\eqref{E:expansion_function_nonconnected} into \eqref{E:chi_MoM} and use the regularized lattice Green's function given by \cite{silveirinha_metamaterial_2007}
\begin{align}\label{E:lattice_Green_function}
\Phi_{p0}(\vec{r}|\vec{r}';\omega,\vec{k})= \frac{1}{V_\text{cell}} \sum_{\vec{J}\neq0} \frac{\e{i\vec{k}_\vec{J}\cdot(\vec{r}-\vec{r}')}}{\vec{k}_\vec{J}\cdot \vec{k}_\vec{J}-\frac{\omega^2}{c^2}} \approx  \frac{1}{V_\text{cell}} \sum_{\vec{J}\neq0} \frac{\e{i\vec{k}_\vec{J}\cdot(\vec{r}-\vec{r}')}}{\vec{k}_\vec{J}^0\cdot \vec{k}_\vec{J}^0 },
\end{align}
where $\vec{k}_\vec{J}=\vec{k} + \vec{k}_\vec{J}^0$ with $\vec{k}_\vec{J}^0=j_1\vec{b}_1+j_2\vec{b}_2+j_3\vec{b}_3$ and the
$\vec{b}_i$s are the reciprocal lattice primitive vectors.
The second identity is valid in the long-wavelength limit, $\omega/c\ll\pi/a$ and $|\vec{k}|\ll \pi/a$ \cite{silveirinha_artificial_2009}.
%
%
After straightforward calculations it is found that:
\begin{align}\label{E:chi_non_connected}
 \chi_{m,n}
  &= \frac{1}{a}   \left[  k_m    k_n   - \left(\frac{\omega}{c}\right)^2    \delta_{m,n}\right] \frac{1}{\beta_{m,n}^2},
\end{align}
where $k_i=\vec{k} \cdot \hat{\vec{u}}_i$ and $\beta_{m,n}$ is a
quantity that depends only on the geometry of the system, and is
given by
%
\begin{align}
 \frac{1}{\beta_{m,n}^2} &= \sum_{\substack{j_n=0\\j_m=0\\ \{j_1,j_2,j_3\}\neq \{0,0,0\}}}   \frac{      \left[ J_0\left(   \left|\vec{k}_\vec{J}^0\right| R  \right)\right]^2  }{\left|\vec{k}_\vec{J}^0\right|^2 } \e{i \vec{k}_\vec{J}^0 \cdot (\vec{r}_m-\vec{r}_n)},
\end{align}
where $\vec{r}_n$ denotes the center of the $n$th wire in the unit
cell (the $n$th wire translated by $-\vec{r}_n$ is centered at the
origin) and $J_0$ is the Bessel function of 1st kind and 0th order. For $m=n$,
$\beta_{m,m}=\beta_p$ is the plasma wavenumber for an array of
parallel PEC wires mentioned in the last subsection
\cite{silveirinha_artificial_2009}.

For $m \neq n$ the parameter $\beta_{m,n}$ is given by a simple
series with an oscillating generic term due to the nonzero complex
exponential coefficient. In contrast, form $m = n$ the parameter
$\beta_{m,n}$ is determined by a double series with the generic term
of summation strictly positive. Due to this reason, one has $ \left|
\frac{1}{\beta_{m,n}^2} \right| \ll \frac{1}{\beta_p^2}$ for $m\neq
n$. The approximation is better for a larger physical distance
between the two sub-lattices, as for a larger distance the complex
exponential will oscillate faster. Thus, the off-diagonal terms of
$\left[\chi_{m,n}\right]$ can be dropped, and with this
approximation the inverse matrix elements are given by:
\begin{align}
 \chi^{m,n}
  &\approx
  \begin{cases}
  \frac{1}{ \chi_{n,n}} \quad  &m= n \\
    0 \quad  &m\neq n
  \end{cases}.
\end{align}
Substituting this expression into \eqref{E:eps_ef_non_connected_1}
it is found that the dielectric function of the double wire medium
is
\begin{align}  \label{E:eps_ef_crossed_wire}
 \frac{\db{\eps}_\text{ef}}{\eps_0}(\omega,\vec{k})  &= \db{\vec{I}} +   \sum_{i}  \frac{\beta_{p}^2}{\left[  k_i^2   - \left(\frac{\omega}{c}\right)^2   \right]}   \hat{\vec{u}}_i    \otimes    \hat{\vec{u}}_i.
\end{align}
This result agrees with the nonlocal effective permittivity for
perfect electric conducting wires derived in \cite{silveirinha_homogenization_2005} using a
slightly different approach. 
Similar to the uniaxial wire medium, the effective permittivity of
the double wire mesh is strongly spatially dispersive. Remarkably,
each wire array contributes independently to the permittivity
function such that ${\overline{\overline \varepsilon }
_{{\rm{ef}}}}/{\varepsilon _0} = \overline{\overline {\bf{I}}}  +
\sum\limits_{n = 1,2} {\left( {{{\overline{\overline \varepsilon }
}_n}/{\varepsilon _0} - \overline{\overline {\bf{I}}} } \right)}$,
where ${{{\overline{\overline \varepsilon } }_n}}$ is the
permittivity of the $n$th wire array alone.

The above derivation can be readily extended to plasmonic wires with
a finite conductivity \cite{silveirinha_nonresonant_2008} and to
triple non-connected wire arrays \cite{silveirinha_homogenization_2005}.
Furthermore, the proof can also be generalized to the case where the
wire arrays are not perpendicular \cite{morgado_negative_2012}. Also in this case, with similar
approximations, one finds that each wire array contributes
independently to the permittivity function.

\subsubsection{3D connected wire mesh}\label{sec:3D_connected_wire_mesh}

The strong spatial dispersion characteristic of nonconnected wire
arrays can be tamed by connecting the metallic wires, so that
effectively the structure is formed by a single piece of conductor \cite{silveirinha_homogenization_2005,demetriadou_taming_2008}.
%
Here, we illustrate this by considering a 3D connected wire mesh
formed by three orthogonal and connected sets of wires as
represented in Fig. \ref{fig:different_wire_meshes}(c).
\\
In this system, because of the discontinuity of the induced current
at the wire junctions, a single expansion function per wire is not
enough to correctly homogenize the electromagnetic response.
Instead, it can be shown that five expansion functions
$\vec{w}_{n,\vec{k}}$ are required to obtain an approximate analytic
expression of the effective permittivity
\cite{silveirinha_artificial_2009,silveirinha_homogenization_2005}.
%
Relying on an approach similar to that of the previous subsection
(the details can be found in \cite{silveirinha_artificial_2009}), it
can be shown that the effective permittivity of this metamaterial is
\begin{align} \label{E:eps_wire_mesh}
\frac{\db{\eps}_\text{ef}}{\eps_0}(\omega,\vec{k})=\eps_t(\omega)\left( \db{\vec{I}} - \frac{\vec{k}\vec{k}}{k^2} \right) + \eps_l(\omega,k) \frac{\vec{k}\vec{k}}{k^2},
\end{align}
where the transverse and longitudinal components are given respectively by
\begin{align}
 \eps_t(\omega)&=1   + \frac{1}{\frac{1}{f_V(\eps_m-1)}-\frac{\omega^2}{\beta_p^2c^2}  }, \\
 \eps_l(\omega,k)&=1   + \frac{1}{\frac{k^2}{l_0\beta_p^2}+\left(\frac{1}{f_V(\eps_m-1)}-\frac{\omega^2}{\beta_p^2c^2} \right) }.
\end{align}
In the above, $l_0=\frac{3}{1+2\beta_p^2/\beta_1^2}$ and $\beta_1$
is a constant (with unities of wave number) that depends solely on
the geometry of the structured material (see
\cite{silveirinha_artificial_2009} for more
details).

Remarkably, the 3D connected wire medium has a homogenized response
equivalent to that of a plasma described by the hydrodynamic model \cite{bittencourt_fundamentals_2004}.
%
%
In particular, the response to transverse waves (with electric field
perpendicular to the wave vector) is described by the
$\bf{k}$-independent transverse permittivity $\eps_t$. However, the
3D connected wire medium remains spatially dispersive. The reason is
that the response to longitudinal waves (with electric field
parallel to the wave vector) is described by a
$\bf{k}$-dependent longitudinal permittivity $\eps_l$. The effects
of spatial dispersion are several orders of magnitude stronger than
in metal nanostructures at optics because the parameter $l_0$ is
relatively small ($l_0 \approx 2$). The effects of spatial
dispersion can be further suppressed by loading the wires with metal
plates, which leads to $l_0 \gg 1$ \cite{demetriadou_taming_2008,maslovski_nonlocal_2009}.
%
%

In general, the 3D connected wire mesh supports 3 electromagnetic
modes: a longitudinal and two transverse plane waves. Propagation is
only feasible above the effective plasma frequency. Thus, for long
wavelengths the 3D connected wire mesh is completely opaque to
radiation. For further details about the electrodynamics of the
connected wire medium, the reader is referred to
\cite{silveirinha_artificial_2009}.

\subsection{Quasi-static model} \label{sec:quasi_static_model}
The nonlocal response of wire metamaterials can be explained by a
quasi-static model developed in \cite{maslovski_nonlocal_2009}. In
the quasi-static model the macroscopic electromagnetic fields are
coupled to the currents in the wires and to an ``additional
potential". The additional potential may be understood as the
average voltage drop from a given wire to the boundary of the cell
wherein it is contained \cite{maslovski_nonlocal_2009}. Both the
additional potential ($\varphi$) and the current are interpolated as
continuous functions defined in all space. As reported in
\cite{maslovski_generalized_2010,silveirinha_radiation_2012}, the
quasi-static model is particularly useful in problems involving
interfaces, e.g., to obtain ``additional boundary conditions", and
to derive conservation laws \cite{silveirinha_radiation_2012}.  

For the case of the uniaxial wire medium (with wires oriented along
$\hat{\vec{z}}$) of section \ref{sec:uniaxial} the quasi-static
model is determined by:
\begin{subequations}\label{E:local_model_uniaxial}
\begin{align}
\nabla \times \vec{E}&=i\omega \mu_0 \vec{H} \\
\nabla \times \vec{H}&=\frac{I_z}{ a^2 }\hat{\vec{z}}-i\omega \eps_h \vec{E} \\
 \frac{\partial I_z}{\partial z} &=i\omega C \varphi \\
  \frac{\partial\varphi}{\partial z}  &= E_z- (Z_w -i \omega L )I_z
\end{align}
\end{subequations}
where $\vec{E}$ and $\vec{H}$ are the macroscopic electromagnetic
%
%
fields ($\vec{E}=\left\{\vec{e}(\vec{r})\right\}_\text{av}$ and $\vec{H}=\left\{\vec{b}(\vec{r})\right\}_\text{av}/\mu_0$), $E_z=\vec{E}\cdot \hat{\vec{z}}$, $\eps_h$ is the
permittivity of the host medium  and $C$, $L$ and $Z_w$ are the
capacitance, inductance and self-impedance of a wire per unit of
length, respectively, defined as in \cite{maslovski_nonlocal_2009}.
As seen, in this theory the macroscopic Maxwell equations are
coupled to a set of differential equations governing the dynamics of
the internal degrees of freedom of the medium ($I_z$ and $\varphi$).
The quasi-static model (Eq. \eqref{E:local_model_uniaxial}) fully
describes the physical behavior of the uniaxial wire medium, as it
can be transformed into the nonlocal model \eqref{E:eps_ef_uniaxial}
by expressing  $I_z$ and $\varphi$ in terms of the macroscopic
fields \cite{maslovski_nonlocal_2009}.

Importantly, the quasi-static model is $local$ as it corresponds to
a standard partial-differential system. The differential operators
act on the 8-component state vector $\left(
{{\bf{E}},{\bf{H}},\varphi ,{I_z}} \right)$. The nonlocality of the
electromagnetic response is a consequence of the fact that $I_z$ and
$\varphi$ are coupled to each other through a space differential
operator ($\partial/\partial z$), different from conventional local
media where the internal degrees of freedom are coupled through time
differential operators ($\partial/\partial t$).

Finally, it is worth mentioning that the quasi-static model is not
restricted to the description of the uniaxial wire medium, as it can
be extended to more complex connected and nonconnected wire medium
topologies \cite{maslovski_nonlocal_2009}.

\subsection{Poynting vector} \label{sec:Poynting_vector}

In spatially-dispersive media, the energy density flux is not given
by the standard textbook formula of the Poynting vector $\vec{E}
\times \vec{H}$
\cite{landau_electrodynamics_1984,agranovich_crystal_1984,
silveirinha_poynting_2009,costa_poynting_2011}. For the case of
lossless materials characterized by a nonlocal dielectric function
the (time-averaged) Poynting vector must instead be calculated
using:
\begin{align}\label{E:poynting_vector_nonlocal}
\vec{S}_\text{av}\cdot \hat{\vec{l}}= \frac{1}{2} \mathrm{Re}\left\{
\left(\vec{E} \times \frac{\vec{B}^\ast}{\mu_0} \right)\cdot
\hat{\vec{l}} \right\} -  \frac{\omega}{4}  \vec{E}^\ast \cdot
\frac{\partial \db{\eps}_\text{ef}}{\partial k_l}(\omega,\vec{k})
\cdot \vec{E}.
\end{align}
Here,  $\hat{\vec{l}}$ is a generic (real-valued) unit vector. It is
implicit that the spatial dependence is of the form $\e{i
\vec{k}\cdot \vec{r}}$ with $\vec{k}$ real-valued and that the
magnetic response is trivial. The formula can be generalized to a
superposition of plane waves possibly associated with complex-valued
wave vectors
\cite{silveirinha_poynting_2009,silveirinha_additional_2009}.

It was demonstrated in Refs.
\cite{silveirinha_poynting_2009,costa_poynting_2011} that for a
generic dielectric metamaterial, Eq.
\ref{E:poynting_vector_nonlocal} agrees precisely with the
cell-averaged microscopic Poynting vector,
\begin{align}
\vec{S}_\text{av} &=\frac{1}{V_\text{cell}}\int_\Omega \frac{1}{2}
\text{Re}\left( \vec{e} \times \frac{\vec{b}^\ast}{\mu_0} \right)
d^3\vec{r},
\end{align}
provided the effective dielectric function is determined with the
homogenization method of Sect.
\ref{sec:electromagnetic_homogenization}. Therefore, the macroscopic
Poynting vector can be understood as a cell-averaged microscopic
Poynting vector.

Evidently, in wire metamaterials the Poynting vector can be
determined using Eq. \ref{E:poynting_vector_nonlocal}, using the
relevant expression of the nonlocal permittivity in the formula.
However, as previously mentioned, such formalism is only applicable
to plane waves. A more general and useful expression for the
Poynting vector can be obtained using the quasi-static model of
section \ref{sec:quasi_static_model}. Indeed, based on Eq.
\eqref{E:local_model_uniaxial} it is possible to derive a
generalized Poynting theorem, which for the particular case of the
uniaxial wire medium yields the following expression for the
Poynting vector \cite{silveirinha_radiation_2012}:
\begin{align}\label{E:poynting_vector_local}
\vec{S}_\text{av}= \frac{1}{2} \mathrm{Re}\left\{  \vec{E} \times\vec{H}^\ast+
\frac{\varphi I_z^\ast}{a^2} \hat{\vec{z}} \right\}.
\end{align}
As seen, the Poynting vector is written in terms of the macroscopic
electromagnetic fields and of the internal degrees of freedom ($I_z$
and $\varphi$) of the metamaterial. It can be verified that in the
lossless case and for a spatial dependence of the form $\e{i
\vec{k}\cdot \vec{r}}$ with $\vec{k}$ real-valued the above
expression reduces to \eqref{E:poynting_vector_nonlocal}. However,
Eq.\eqref{E:poynting_vector_local} is more general than
\eqref{E:poynting_vector_nonlocal} as it can be applied to arbitrary
electromagnetic field distributions. The stored energy in the wire
metamaterial can also be expressed in terms of the state vector
$\left( {{\bf{E}},{\bf{H}},\varphi ,{I_z}} \right)$, and for more
details the reader is referred to \cite{silveirinha_radiation_2012}.

\subsection{Additional boundary conditions} \label{sec:ABC}

One important consequence of spatial dispersion is that the usual
Maxwellian boundary conditions, i.e., the continuity of the
tangential $\bf{E}$ and $\bf{H}$  fields, are insufficient to solve
wave propagation problems in the presence of interfaces
\cite{agranovich_crystal_1984,nefedov_reflection_2006,silveirinha_additional_2009,yatsenko_higher_2000,nefedov_electromagnetic_2005,pekar_1958,silveirinha_additional_2006,yakovlev_generalized_2011}.
For example, consider a planar interface between two regions: a
standard dielectric and a generic spatially dispersive material
characterized by a nonlocal dielectric function. Suppose that a
plane wave propagating in the dielectric illuminates the
spatially-dispersive material half-space. The standard approach to
find the scattered waves is to expand the electromagnetic fields
into plane waves in the two regions and then to match the fields at
the interfaces by imposing the standard Maxwellian boundary
conditions. In standard dielectrics, there are exactly two
plane-waves associated with an energy flow propagating away from the
interface, i.e., there are only two polarization states per
propagation direction. The potential problem is that in a nonlocal
material the allowed number of polarization states per propagation
direction may be greater than two, i.e., the medium may support
``additional" waves. For example, a uniaxial wire medium typically
supports three independent polarization states
\cite{silveirinha_additional_2006}. Consequently, it is generally
impossible to solve a scattering problem relying only on the
Maxwellian boundary conditions because the number of unknowns
(number of waves that can be excited) is greater than the number of
equations (number of boundary conditions). The problem is
under-determined and additional boundary conditions (ABCs) are
needed. The number of ABCs must be the same as the number of
additional waves.

For wire metamaterials, the ABC requirement is particularly clear
from the quasi-static formulation of Sec.
\ref{sec:quasi_static_model} where it is evident that in a
scattering problem the boundary conditions for the internal degrees
of freedom $\varphi$ and $I_z$ must also be provided
\cite{maslovski_generalized_2010}. Thus, one  needs to specify how
the relevant internal variables behave at the interface.
Unfortunately there is no systematic theory to find the ABCs, and
their derivation must be based on the specific microscopic
properties of the system under consideration. In particular, it is
underlined that the ABCs (which are interface dependent) cannot be
directly obtained from the nonlocal dielectric function, i.e., from
the bulk response.

Here, we restrict our attention to an interface between a wire
metamaterial and a standard dielectric. This situation covers the
important case of an interface between wire media and air, which is
of particular interest for scattering or imaging applications.
Evidently, the microscopic electric currents in the metal wires are
interrupted at the interface. Hence, for a system with $N$
independent wires in the unit cell, it follows that at the
dielectric interface
\begin{align}\label{E:ABC}
 \vec{J}_{\text{av}} \cdot \hat{\vec{u}}_n = 0, \quad n=1, \dots, N
\end{align}
where $\vec{J}_{\text{av}}$ is the cell-averaged microscopic
conduction current and $\hat{\vec{u}}_n$ is the unit vector oriented
along the direction of the $n$-th wire array
\cite{silveirinha_additional_2009}. The vector $\vec{J}_{\text{av}}$
can typically be written in terms of the dielectric function of the
medium \cite{silveirinha_additional_2009}.

In the particular case of a uniaxial wire medium, the ABC in the
quasi-static model assumes the simple and intuitive form $I_z=0$.
This ABC (together with the standard Maxwellian boundary conditions)
can be expressed in terms of the electromagnetic fields as
\cite{silveirinha_additional_2006}:
\begin{align}\label{E:ABC_WM}
{\varepsilon _h}{\left. {\hat{\bf{n}} \cdot {\bf{E}}}
\right|_{{\rm{WM}}}} = {\varepsilon _d}{\left. {\hat{\bf{n}}  \cdot
{\bf{E}}} \right|_{{\rm{diel}}{\rm{.}}}},
\end{align}
where $\hat{\bf{n}}$ is the normal to the interface,
$\varepsilon_h$ is the host medium permittivity and $\varepsilon_d$
is the dielectric permittivity. Note that equation \eqref{E:ABC_WM} is not equivalent to the continuity of the electric displacement vector, since the effective permittivity of the wire medium is different from $\eps_h$. Similar ideas are used to obtain the
ABCs for the case of connected wire arrays
\cite{silveirinha_artificial_2009}, interlaced wire meshes
\cite{hanson_modeling_2012} and for wires terminated with lumped
loads \cite{maslovski_generalized_2010,silveirinha_additional_2009}.

It should be noted that wire metamaterials are amongst the very few examples of structured media for which there is a clear understanding of how to model the nonlocal effects near interfaces \cite{maslovski_generalized_2010,silveirinha_additional_2009,silveirinha_additional_2006,hanson_non-local_2013}. Another example, less well-developed, is the case of quadrupolar metamaterials characterized by weak spatial dispersion \cite{silveirinha_boundary_2014,yaghjian_boundary_2014}.
The general problem of characterizing the  interface response of a generic nonlocal metamaterial is unsolved.


\section{Anomalous refraction and light tunneling with wire metamaterials} \label{sec:anomalous_light_refraction_tunneling} 
To illustrate some of the unusual opportunities created by the spatial
dispersion in wire metamaterials, we review in the next subsections
the effects of anomalous light refraction and anomalous light
tunneling.

\subsection{Anomalous refraction in arrays of non-connected crossed metallic wires} \label{sec:prism} 
\begin{figure*}[!ht]
\centering
\includegraphics[width=.85\linewidth]{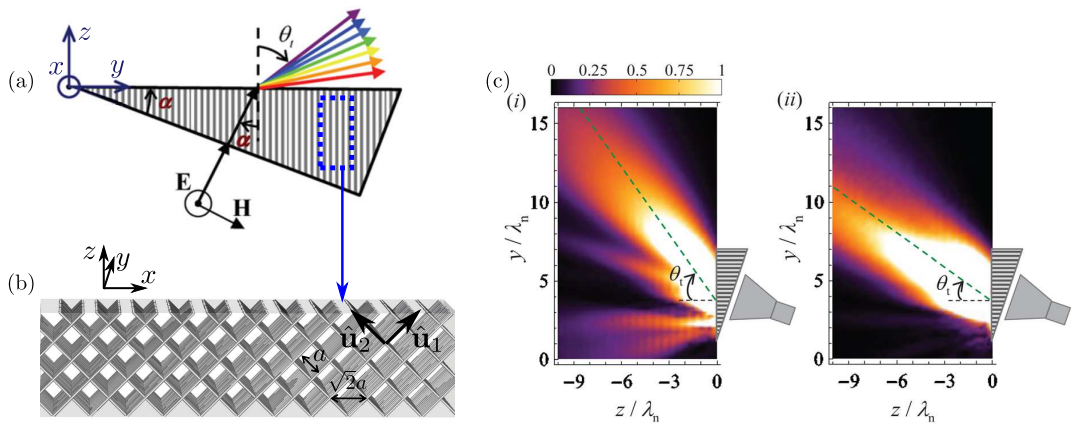}
         \caption{(a) Schematic of the anomalous light refraction in a prism made of non-connected arrays of parallel wires.
(b) Each array of parallel wires is arranged in a square lattice
with lattice constant $a$. The two arrays of wires are mutually
orthogonal and lie in planes parallel to the $x$-axis. The
distance between adjacent perpendicular wires is $a/2$. The metallic
wires are tilted by $\pm45^\circ$ with respect to the $xoy$ plane.
(c) Normalized squared amplitude of the measured electric field for
a prism made of a crossed array of metallic strips for a frequency
of (i) $7.605 \mathrm{GHz}$ (ii) $16.325 \mathrm{GHz}$. A schematic
drawing of the metamaterial prism and horn antenna (at full scale)
is shown. The propagation is towards the left-hand side region. The
green dashed lines represent the refracted beam propagation
direction, whereas the black dashed lines represent the direction
normal to the interface. The spatial coordinates $y$ and $z$ are
normalized to the reference wavelength $\lambda_n=39.71 \mathrm{mm}$
\cite{morgado_reversed_2014}.}
         \label{fig:prism}
\end{figure*}

As noticed in \cite{silveirinha_anomalous_2009}, a remarkable consequence of
spatial dispersion is the possibility to achieve a low-loss and
broadband regime of anomalous light refraction such that, contrary
to what happens in a standard glass prism, longer wavelengths are
more refracted than shorter wavelengths. This effect is forbidden by
Kramers-Kronig relations in transparent and local materials. It may
however occur in a prism made of a double-wire medium formed by
nonconnected wires \cite{silveirinha_anomalous_2009,
morgado_reversed_2014}, see Fig.\ref{fig:prism}.
%
%
%
To understand the physical origin of this effect, we consider the
wave propagation in an unbounded double wire medium made of
perfectly conducting wires lying in the $xoz$ plane and tilted by
$\pm45^\circ$ with respect to the $x$-axis, as represented in
Fig.\ref{fig:prism} (b). For simplicity, we assume that the wave
propagates along  the $z$-direction ($\vec{k}=k_z\hat{\vec{z}}$).
For fields polarized along the $x$-direction the characteristic
equation is
\begin{align} \label{E:characteristic_eq_crossed}
\eps_{xx}(\omega,k_z) \frac{\omega^2}{c^2}= k_z^2,
\end{align}
where $\eps_{xx}=\hat{\vec{x}} \cdot \db{\eps}_\text{ef}\cdot
\hat{\vec{x}}$ is the relevant component of the nonlocal effective
permittivity for this polarization. According to the effective model
\eqref{E:eps_ef_crossed_wire}, $\eps_{xx}$  is given by
\begin{align}
\eps_{xx}(\omega,k_z) = 1 +
\frac{\beta_p^2}{\frac{k_z^2}{2}-\frac{\omega^2}{c^2}}.
\end{align}
%
%
Substituting $\eps_{xx}$ into Eq.
\eqref{E:characteristic_eq_crossed} and solving for $k_z$, it is
found that $k_z=\frac{\omega}{c}n_\text{ef}$, where $n_\text{ef}$ is
the effective refractive index of the double wire medium given by
\cite{silveirinha_anomalous_2009}
\begin{align}
n_\text{ef} = \sqrt{\frac{3}{2}+ \frac{1}{2}\sqrt{1+8\left(\frac{\beta_p c}{\omega}\right)^2}}.
\end{align}
Remarkably, even though the metamaterial is lossless, the refractive
index is a strictly decreasing function of the frequency. This
unique property is only possible due to the spatial dispersion which
makes the permittivity seen by the transverse field (polarized along
$x$) dependent on a perpendicular wave vector component (here
$k_z$). The same effect occurs for other propagation directions in
the $yoz$ plane.

Due to the anomalous permittivity dispersion, a prism made of a
crossed wire mesh can create a reverse rainbow as demonstrated
theoretically in \cite{silveirinha_anomalous_2009}, and
experimentally confirmed at microwave frequencies in
\cite{morgado_reversed_2014}. In the experiment the prism is formed
by a stack of dielectric slabs printed with the $\pm
45^{\circ}$-oriented metallic strips. A sample of the experimental
results is presented in Fig.\ref{fig:prism}(c). As seen, unlike
conventional prisms, in the metamaterial prism the refracted beam
comes out closer to the normal of the output interface when the
frequency is increased. Materials with anomalous light dispersion
may be useful for many applications, e.g., for the compression of
light pulses or for the correction of achromatic aberrations
\cite{costa_achromatic_2012}.

\subsection{Anomalous light tunneling in interlaced wire meshes}\label{sec:interlaced_wire_mesh}

\begin{figure*}[!ht]
\centering
\includegraphics[width=.65\linewidth]{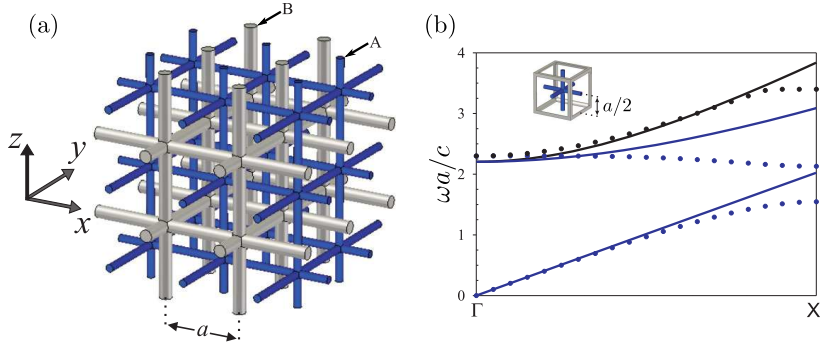}
         \caption{(a) Geometry of the interlaced wire mesh. The wire radii in mesh $A$ and $B$ are $r_A$ and $r_B$, respectively. The wires of each network are spaced by a distance $a$ along the coordinate axes. The distance between the two non-connected networks is $a/2$. (b) Band diagram of the electromagnetic modes along the direction $\Gamma X$. Solid lines: analytical model; discrete symbols: full wave simulations. The inset shows the cubic unit cell of the structure. The wires are PEC and are embedded in a dielectric with permittivity $\eps_h = 1$; the radii of the wires are $r_A= 0.001a$ and $ r_B = 0.05a$. Reprinted with permission \cite{latioui_light_2017}.}
         \label{fig:interlaced_wire_mesh}
\end{figure*}

Here, we consider a metamaterial formed by two interlaced 3D
connected wire meshes (mesh $A$ and $B$) separated by half-lattice
period $a$, see Fig.\ref{fig:interlaced_wire_mesh} (a)
\cite{hanson_modeling_2012,shin_three-dimensional_2007}. In what
follows, we characterize the effective response of this ``interlaced
wire medium" and discuss a counter-intuitive tunneling effect rooted
in the spatially dispersive response of the metamaterial.

Consider the general problem of homogenization of a metamaterial
formed by two networks of inclusions $A$ and $B$. In general, to
find the effective response it is crucial to take into account the
complex near-field interactions between the different types of
scatterers. However, when the scatterers are physically distant in
the unit cell it may be a good approximation to consider that each
scatterer behaves as a ``macroscopic source'' from the point of view
of the other scatterer. Essentially, this approximation is good when
only the smooth (slowly varying) part of the fields radiated by one
of the scatterers influences the currents on the other scatterer. It
can be formally shown that in these conditions each component of the
metamaterial contributes independently to the dielectric function
such that \cite{hanson_modeling_2012,fernandes_fano_2013}:
\begin{align} \label{E:interlacedWM}
\db{\eps}_\text{ef}=\db{\eps}_\text{ef}^A+\db{\eps}_\text{ef}^B-
\eps_h \db{\vec{I}},
\end{align}
where $\db{\eps}_\text{ef}^i$ with $i=A,B$ is the nonlocal effective
permittivity of the metamaterial formed only by the $i$th network of
inclusions.

From the results of Sect. \ref{sec:non_connected_wire_mesh}, it is
readily recognized that a double-wired mesh of nonconnected wires
provides a nontrivial example of a system in which the different
types of scatterers interact with one another as ``macroscopic
sources''. Interestingly, it turns out that the interlaced wire mesh
of Fig.\ref{fig:interlaced_wire_mesh}(a) has the same property when
the two 3D wire meshes are separated by the maximal possible
distance ($a/2$) \cite{hanson_modeling_2012, latioui_light_2017}.
For the interlaced wire mesh $\db{\eps}_\text{ef}^i$ is the nonlocal
effective permittivity of the (isolated) $i$th wire mesh given by
Eq.\eqref{E:eps_wire_mesh}.

Intuitively, the interlaced wire mesh should be opaque to radiation
for frequencies below a certain effective plasma frequency.
Surprisingly, that is not the case and it turns out that the
metamaterial supports a longitudinal-type mode at
arbitrary low-frequencies, as illustrated in the band diagram in
Fig.\ref{fig:interlaced_wire_mesh}(b)
\cite{hanson_modeling_2012,shin_three-dimensional_2007,latioui_light_2017}.
This feature contrasts sharply with the properties of the individual
3D wire meshes, which do not support electromagnetic propagation for
long wavelengths.

\begin{figure*}[!ht]
\centering
\includegraphics[width=\linewidth]{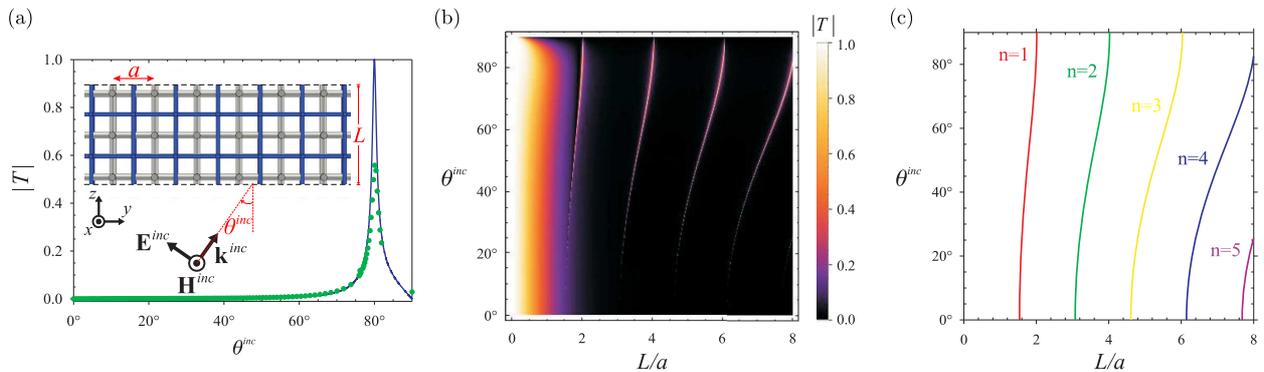}
         \caption{(a)  Amplitude of the transmission coefficient as a function of the incidence angle $\theta^\text{inc}$ for the normalized frequency $\omega a/c = 1.32$ and normalized thickness $L/a \approx 6$. The remaining structural parameters are as in Fig. \ref{fig:interlaced_wire_mesh}. The inset shows the geometry of the problem. The solid lines represent the analytical results, and the discrete symbols represent the full wave simulations results. (b) Density plot of the transmission coefficient amplitude as a function of the normalized thickness $L/a$ and of the incidence angle $\theta^\text{inc}$ at the fixed frequency of $\omega a/c = 1.32$. (c) Incidence angle $\theta^\text{inc}$ as a function of $L/a$ for the $n$th ($n=1,2,\dots$) Fabry-Pérot resonance of the propagating longitudinal mode at $\omega a/c = 1.32$.}
         \label{fig:interlaced_wire_mesh_result}
\end{figure*}

Remarkably, the low-frequency mode can originate a rather
counter-intuitive tunneling effect. To illustrate this, we consider
an interlaced wire mesh slab of finite length $L$ (see the inset of
Fig.\ref{fig:interlaced_wire_mesh_result}(a)). Using the effective
permittivity model \eqref{E:interlacedWM} and suitable ABCs, it is
possible to find the transmission coefficient $\left|T\right|$ of
the slab \cite{latioui_light_2017}.
Strikingly, as shown in Fig.
\ref{fig:interlaced_wire_mesh_result}(a), provided the wire radii
are different ($r_A\neq r_B$) an incoming plane wave can tunnel
through the metamaterial slab for large incidence angles. This
transmission anomaly is due to a Fabry-P\'erot resonance of
the low-frequency longitudinal mode of the metamaterial. At the
resonance  the longitudinal wave vector satisfies the condition $k_z
L= n \pi$, with $n=1,2,\dots$ (see
Figs.\ref{fig:interlaced_wire_mesh_result} (b) and (c)).

The physical origin of the tunneling anomaly is a Fano-type
resonance \cite{fernandes_fano_2013} that occurs when $r_A\neq r_B$
and enables the cancellation due to destructive interferences of the  scattering by the two
subcomponents of the interlaced wire mesh. This metamaterial
structure may be useful for angle-dependent filtering and sensing.
For a detailed discussion of the physical properties of the
interlaced wire mesh, the reader is referred to
\cite{latioui_light_2017}.

\section{Conclusions}
We presented an overview of a first principles homogenization
approach based on an effective Hamiltonian that describes exactly
the time evolution of the wave packet envelope when the initial
state is less localized than the lattice period. The effective
Hamiltonian determines completely the band diagram of the
time-stationary states of the periodic system. The homogenization
formalism can be applied to a wide range of physical systems. Its
specific implementation for the case of nonmagnetic periodic
electromagnetic metamaterials was detailed.

In particular, we focused our attention in the homogenization of
wire metamaterials with diverse topologies. These structures are
typically characterized by a strong nonlocal response in the long
wavelength limit. In wire metamaterials formed by two or more non-connected networks, each metal network may contribute almost independently to the permittivity function. We underlined the nontrivial implications of the
spatially dispersive response in different contexts, e.g., the
emergence of additional waves, additional boundary conditions and a
non-standard definition of the Poynting vector. Finally, we
illustrated the richness of the physics of the wave propagation in
wire medium, showing that it can lead to a counter-intuitive
tunneling effect and anomalous frequency dispersion.

\end{document}